\documentclass[seceq]{ptptex}

\usepackage{graphicx}
\usepackage{amsmath}

\notypesetlogo                       

\title{
A Bayesian Approach to QCD Sum Rules%
}

\author{
Philipp \textsc{Gubler}\footnote{E-mail: phil@th.phys.titech.ac.jp} and Makoto \textsc{Oka}
}

\inst{
Department of Physics, Tokyo Institute of Technology, \\ Tokyo 152-8551, Japan 
}

\abst{
QCD sum rules are analyzed with the help of the Maximum Entropy Method. 
We develop a new technique based 
on the Bayesion inference theory, which 
allows us to directly obtain the spectral function 
of a given correlator from the results of the operator product expansion 
given in the deep euclidean 4-momentum region. The most important advantage of this 
approach is that one does not have to make any a priori assumptions about the functional 
form of the spectral function, such 
as the ``pole + continuum" ansatz that has been widely used in QCD sum rule studies, 
but only needs to specify the asymptotic values of the spectral function at high and low energies 
as an input. 
As a first test of the applicability of this method, we have analyzed the sum rules of the $\rho$-meson, 
a case where the sum rules are known to work well. Our results show a clear peak structure in the region 
of the experimental mass of the $\rho$-meson. We thus demonstrate that the Maximum Entropy Method is 
successfully applied and that it is an efficient tool in the analysis of QCD sum rules. 
}

\begin{document}

\maketitle

\section{Introduction}
The technique of QCD sum rules is well known for its ability to reproduce various properties of 
hadrons \cite{Shifman,Reinders}. Using dispersion relations, this method connects perturbative 
and nonperturbative sectors of QCD, and therefore, allows one to describe inherently 
nonperturbative objects such as hadrons by the operator product expansion (OPE), which is 
essentially a perturbative procedure. The higher-order terms of the OPE contain condensates 
of various operators, which incorporate information on the QCD vacuum. Hence, QCD sum rules 
also provide us with nontrivial relations between the properties of hadrons and the QCD 
vacuum.

Since the early days of the development of QCD sum rules, the range of applications of this method 
has been constantly expanding, which has helped to explain many aspects of the behavior of hadrons. 
Nevertheless, QCD sum rules have always been subject to (justified) criticism. One part of this 
criticism is of mainly technical nature, pointing out that the analysis of QCD sum rules 
often is not done with the necessary rigor, namely, that 
the OPE convergence and/or the pole dominance condition are not properly taken into account. 
Many of the recent works that followed the claimed discovery of the pentaquark $\Theta^+(1540)$ 
are examples of such a lack of rigor. 
Nonetheless, these technical problems can be overcome if the analysis is done 
carefully enough \cite{Gubler1,Gubler2}. 

The second part of the criticism against QCD sum rules is 
more essential. It is concerned with the 
ansatz taken to parametrize the spectral function. For instance, it is common to assume 
the ``pole + continuum" functional form, where the pole represents the hadron in 
question and the continuum stands for the excited and scattering states. 
While this ansatz may be justified in 
cases where the low-energy part of the spectral function is dominated by a single 
pole and the continuum states become important only at higher energies (the $\rho$-meson 
channel for instance is such a case), it is not at all clear if it is also valid in other cases. 
For example, as shown in Ref. \citen{Kojo}, where the $\sigma$-meson channel was 
investigated using QCD sum rules, it can be difficult to distinguish the continuum spectrum from 
a broad resonance, because they lead to a similar behavior of the ``pole" mass and residue 
as functions of the Borel mass. 
Moreover, using the ``pole + continuum" ansatz, the outcome of the analysis usually 
depends on unphysical parameters such as the Borel mass or the threshold parameter, and 
it is not always a trivial matter to determine these parameters in a consistent way. 
After all, the ansatz of parametrizing the spectral function makes a full 
error estimation impossible in QCD sum rules. 

As a possible solution to these problems, we propose to analyze QCD sum rules with the 
help of the Maximum Entropy Method (MEM). This method has already been applied to 
Monte-Carlo studies in both condensed matter physics \cite{Jarrel} and lattice 
QCD \cite{Nakahara,Asakawa}, and has applications in many other areas \cite{Wu}. 
It makes use of Bayes' theorem of probability theory, which helps to 
incorporate known properties of the spectral function such as positivity 
and asymptotic values into 
the analysis and finally makes it possible to obtain the most probable 
spectral function without having to introduce any additional a priori assumptions about 
its explicit form. 
It even allows us to estimate the error of the obtained spectral function. 
Therefore, using this approach, it should in principle be 
possible to study the spectral function of any channel, including those for which 
the ``pole + continuum" assumption is not appropriate.

However, as a first step it is indispensable to check whether QCD sum rules are a 
suitable target for MEM and if 
it is possible to obtain any meaningful information on the spectral function by 
this method. To provide an answer to these questions is the main object of this paper. 
To carry out this check, we have chosen to investigate the sum rule of the $\rho$-meson. 
This channel is one of the first subjects that have been studied in QCD sum rules and 
it is fair to say that it is the channel where this method has so far seen its 
most impressive success. As mentioned earlier, it is a case where the 
``pole + continuum" ansatz works well and we thus do not expect to gain anything 
really new from this analysis. Nevertheless, apart from the aspect of testing 
the applicability of our new approach, we believe that it is worth examining this 
channel once more, as MEM also provides a new viewpoint of looking at various 
aspects of this particular sum rule.

The paper is organized as follows. In \S2, the formalism of both QCD sum rules 
and MEM are briefly recapitulated. Then, in \S3, we show the findings of 
a detailed MEM analysis using mock data with realistic errors. This is followed by 
the results of the investigation of the actual sum rule of the $\rho$-meson 
channel in \S4. Finally, the summary and conclusions are given in \S5.

\section{Formalism}
\subsection{Basics of QCD sum rules}
The essential idea of the QCD sum rule approach \cite{Shifman,Reinders} is to make use of the analytic properties of the 
two-point function of a general operator $J(x)$:
\begin{equation}
\Pi(q^2) = i \int d^{4}x e^{iqx} \langle0|T[J(x) J^{\dagger}(0)]|0\rangle.  \label{eq:cor1}
\end{equation}
The analytic properties of this expression allow one to write down a 
dispersion relation, which connects the imaginary part of $\Pi(q^2)$ with 
its values in the deep euclidean region, where $q$ is large and spacelike 
($-q^2 \to \infty$). This is the region, where it is possible to systematically 
carry out the operator product expansion. The dispersion relation reads as: 
\begin{equation}
\Pi(q^2) = \frac{1}{\pi}\displaystyle \int^{\infty}_{0} ds \frac{\mathrm{Im} \Pi(s)}{s-q^2} 
+(\text{subtraction terms}). \label{eq:disp} 
\end{equation}
Here, the ``subtraction terms" are polynomials in $q^2$, which have to be introduced to 
cancel divergent terms in the integral on the right-hand side of Eq. (\ref{eq:disp}). When 
the Borel transformation is applied, these terms vanish and we therefore do not consider 
them any longer 
in the following. Using the Borel transformation, defined below, also has the advantage of improving 
the convergence of the operator product expansion. 
\begin{equation}
\widehat{L_M} \equiv \lim_{\genfrac{}{}{0pt}{}{-q^2,n \to \infty,}{
-q^2/n=M^2}}
\frac{(-q^2)^n}{(n-1)!} \Bigg(\frac{d}{dq^2}\Bigg)^n .
\label{eq:Boreltrans1}
\end{equation}
After having calculated the OPE of $\Pi(q^2)$ in the deep euclidean region, 
the Borel transformation is applied to both sides of 
Eq. (\ref{eq:disp}), giving the following expression for 
$G_{OPE}(M) \equiv \widehat{L_M} [\Pi^{OPE}(q^2)]$, which 
depends only on $M$, the Borel mass:
\begin{equation}
\begin{split}
G_{OPE}(M) &= \frac{1}{\pi M^2} \displaystyle \int_0^{\infty}  ds e^{-s/M^2} \mathrm{Im} \Pi(s) \\
&= \frac{2}{M^2} \displaystyle \int_0^{\infty} d \omega   e^{-\omega^2/M^2}  \omega \rho(\omega).
\end{split}
\label{eq:fsumr}
\end{equation}
Here, $s=\omega^2$ and $\mathrm{Im} \Pi(s) \equiv \pi \rho(\omega)$ were used. 
At this point, one usually has to make some assumptions about the explicit form of the spectral function 
$\rho(\omega)$. The most popular choice is the ``pole + continuum" assumption, where one 
introduces a $\delta$-function for the ground state and parametrizes the higher-energy continuum states 
in terms of the expression 
obtained from the OPE, multiplied by a step function:
\begin{equation}
\mathrm{Im}\Pi(s) = \pi|\lambda|^2\delta(s - m^2) + \theta(s - s_{th})\mathrm{Im}\Pi^{OPE}(s). \label{eq:pole}
\end{equation}
Here, $|\lambda|^2$ is the residue of the ground-state pole and $s_{\mathrm{th}}$ determines the scale, above which the 
spectral function can be reliably described using the perturbative OPE expression. It is important to note that 
$s_{\mathrm{th}}$ in many cases does not directly correspond to any physical threshold, which often makes its interpretation 
rather difficult.

We will in the present study follow a different path and do not use Eq. (\ref{eq:pole}) or any other 
parameterizations of the spectral function. Instead, we will go back to the more general Eq. (\ref{eq:fsumr}) and 
employ the Maximum Entropy Method to directly extract the spectral function from this equation and the known 
properties of the spectral function at high and low energies. 
This procedure 
reduces the amount of assumptions that have to be made in the analysis.
  
\subsection{Basics of the Maximum Entropy Method (MEM)}
In this subsection, the essential steps of the MEM procedure are reviewed in brief. For more details, 
see Ref. \citen{Jarrel}, where applications in condensed matter physics are discussed, or 
Refs. \citen{Nakahara} and \citen{Asakawa} for 
the implementation of this method to lattice QCD. For the application of MEM to Dyson-Schwinger studies, see 
also Ref. \citen{Nickel}.

The basic problem that is solved with the help of MEM is the following. Suppose one 
wants to calculate some function $\rho(\omega)$, but has only information about 
an integral of $\rho(\omega)$ multiplied by a kernel $K(M,\omega)$:
\begin{equation}
G_{\rho}(M) = \displaystyle \int_0^{\infty} d\omega K(M,\omega) \rho(\omega). \label{eq:basicproblem}
\end{equation}
For QCD sum rules, this equation corresponds to Eq. (\ref{eq:fsumr}) with 
\begin{equation}
K(M,\omega)=\frac{2\omega}{M^2} e^{-\omega^2/M^2} \label{eq:kernel}
\end{equation} 
and $G_{OPE}(M)$ instead of $G_{\rho}(M)$. 
When $G_{OPE}(M)$ is known 
only with limited precision or is only calculable in a limited range of $M$, the problem 
of obtaining $\rho(\omega)$ from $G_{OPE}(M)$ is ill-posed and will not be analytically solvable. 

The idea of the MEM approach is now to use Bayes' theorem, by which additional information 
about $\rho(\omega)$ such as positivity and/or its asymptotic behavior at small or 
large energies can be added to 
the analysis in a systematic way and by which one can finally deduce the most probable form 
of $\rho(\omega)$. Bayes' theorem can be written as 
\begin{equation}
P[\rho|GH] = \frac{P[G|\rho H] P[\rho|H]}{P[G|H]}, \label{eq:bayes}
\end{equation} 
where prior knowledge about $\rho(\omega)$ is denoted as $H$, and $P[\rho|GH]$ stands for 
the conditional probability of $\rho(\omega)$ given $G_{OPE}(M)$ and $H$. Maximizing this functional 
will give the most probable $\rho(\omega)$. $P[G|\rho H]$ is the ``likelihood 
function" and $P[\rho|H]$ the ``prior probability". Ignoring the prior probability and 
maximizing only the likelihood function corresponds to the ordinary $\chi^2$-fitting. 
The constant term $P[G|H]$ in the denominator is just a normalization constant and 
can be dropped as it does not depend on $\rho(\omega)$. 

\subsubsection{The likelihood function and the prior probability}
We will now briefly discuss the likelihood function and the prior probability one after the 
other. Considering first the likelihood function, 
it is assumed that the function $G_{OPE}(M)$ is 
distributed according to uncorrelated Gaussian distributions. 
For our analysis of QCD sum rules, we will 
numerically generate uncorrelated Gaussianly distributed values 
for $G_{OPE}(M)$, 
which satisfy this assumption. 
We can therefore write for $P[G|\rho H]$: 
\begin{equation}
\begin{split}
P[G|\rho H] =& e^{-L[\rho]}, \\
L[\rho] =& \frac{1}{2(M_{\mathrm{max}} -M_{\mathrm{min}})}  
\displaystyle \int_{M_{\mathrm{min}}}^{M_{\mathrm{max}}} 
dM \frac{ \bigl[G_{OPE}(M) - G_{\rho}(M) \bigr]^2}{\sigma^2(M)}. 
\end{split}
\label{eq:likelihood}
\end{equation}
Here, $G_{OPE}(M)$ is obtained from the OPE of the two-point function and $G_{\rho}(M)$ is defined as in Eq. (\ref{eq:basicproblem}) 
and, hence, implicitly depends on $\rho(\omega)$. 
$\sigma(M)$ stands for the uncertaintity of $G_{OPE}(M)$ at the corresponding value of the Borel mass. 
In practice, we will discretize both the integrals in Eqs. (\ref{eq:basicproblem}) and 
(\ref{eq:likelihood}) and take $N_{\omega}$ data points for $\rho(\omega)$ in the range 
$\omega = 0 \sim \omega_{\mathrm{max}} = 6.0$ GeV and $N_M$ data points 
for $G_{OPE}(M)$ and $G_{\rho}(M)$ in the range from $M_{\mathrm{min}}$ to  $M_{\mathrm{max}}$.

The prior probability parametrizes the prior knowledge of $\rho(\omega)$ such as positivity 
and the values at the limiting points, and is given by 
the Shannon-Jaynes entropy $S[\rho]$:
\begin{equation}
\begin{split}
P[\rho|H] &= e^{\alpha S[\rho]}, \\
S[\rho] &= \displaystyle \int_0^{\infty} d\omega \Bigr[ \rho(\omega) - m(\omega) - \rho(\omega)\log \Bigl( \frac{\rho(\omega)}{m(\omega)} \Bigr) \Bigl].
\end{split}
\label{eq:prior}
\end{equation}
For the derivation of this expression, see for instance Ref. \citen{Asakawa}. It can be either derived from the law of large numbers or 
axiomatically constructed from requirements such as locality, coordinate invariance, system independence and scaling. For our 
purposes, it is important to note that this functional gives the most unbiased probability for the positive function 
$\rho(\omega)$. The scaling factor $\alpha$ that is newly introduced in Eq. (\ref{eq:prior}) will be integrated out in a later step 
of the MEM procedure. The function $m(\omega)$, which is also introduced in Eq. (\ref{eq:prior}), is the so-called ``default model". 
In the case of no available data $G_{OPE}(M)$, the MEM procedure will just give $m(\omega)$ for $\rho(\omega)$ because this function maximizes $P[\rho|H]$. 
The default model is often taken to be a constant, but one can also use it to incorporate known information about $\rho(\omega)$ into the 
calculation. In our QCD sum rule analysis, we will use $m(\omega)$ to fix the value of $\rho(\omega)$ at both very low and large 
energies. As for the other integrals above, we will discretize the integral in Eq. (\ref{eq:prior}) and approximate it with 
the sum of $N_{\omega}$ data points in the actual calculation.

\subsubsection{The numerical analysis}
Assembling the results from above, we obtain the final form for the probability $P[\rho|GH]$:
\begin{equation}
\begin{split}
P[\rho|GH] \propto & P[G|\rho H] P[\rho|H]  \\
=& e^{Q[\rho]}, \\
Q[\rho] \equiv & \alpha S[\rho] - L[\rho].
\end{split}
\label{eq:finalprob}
\end{equation}
It is now merely a numerical problem to obtain the form of $\rho(\omega)$ that maximizes $Q[\rho]$ and, therefore, is the 
most probable $\rho(\omega)$ given $G_{OPE}(M)$ and $H$. As shown for instance in Ref. \citen{Asakawa}, it can be proven that 
the maximum of $Q[\rho]$ is unique if it exists and, therefore, the problem of local minima does not occur. For the 
numerical determination of the maximum of $Q[\rho]$, we use the Bryan algorithm \cite{Bryan}, which uses the singular-value 
decomposition to reduce the dimension of the configuration space of $\rho(\omega)$ and, therefore, largely shortens 
the necessary calculation time. 
We have also introduced some slight modifications to the algorithm, 
the most important one being that once a maximum is reached, we add to $\rho(\omega)$ a 
randomly generated small function and let the search start once again. If the result of the second search agrees with 
the first one within the requested accuracy, it is taken as the final result. If not, we add to $\rho(\omega)$ another 
randomly generated function and start the whole process from the beginning. We found that this convergence criterion 
stabilizes the algorithm considerably. 
   
Once a $\rho_{\alpha}(\omega)$ that maximizes $Q[\rho]$ for a fixed value of $\alpha$ is found, this parameter is integrated out by 
averaging $\rho(\omega)$ over a range of values of $\alpha$ and assuming that $P[\rho|GH]$ is sharply peaked around 
its maximum $P[\rho_{\alpha}|GH]$: 
\begin{equation}
\begin{split}
\rho_{\mathrm{out}}(\omega) &= \displaystyle \int [d \rho] \displaystyle \int d \alpha \rho(\omega) P[\rho|GH] P[\alpha |GH]  \\
& \simeq \displaystyle \int d \alpha \rho_{\alpha}(\omega) P[\alpha |GH].
\end{split}
\label{eq:average}
\end{equation}
This $\rho_{\mathrm{out}}(\omega)$ is our final result. 
To estimate the above integral, we once again make use of Bayes' theorem to obtain $P[\alpha |GH]$ 
(for the derivation of this expression, see again Ref. \citen{Asakawa}):
\begin{equation}
P[\alpha |GH] \propto P[\alpha |H] \exp \Bigr[ \frac{1}{2} \displaystyle \sum_k \log \frac{\alpha}{\alpha + \lambda_k} 
+ Q[\rho_{\alpha}] \Bigl].
\label{eq:weight}
\end{equation}
Here, $\lambda_k$ represents the eigenvalues of the matrix 
\begin{equation}
\Lambda_{ij} = \sqrt{\rho_i} \frac{\partial^2 L}{\partial \rho_i \partial \rho_j} \sqrt{\rho_j} \bigg|_{\rho=\rho_{\alpha}}, 
\end{equation}
where $\rho_i$ stands for the discretized data points of $\rho(\omega)$: $\rho_i \equiv \rho(\omega_i) \Delta \omega$, 
with $\Delta \omega \equiv \frac{\omega_{\text{max}}}{N_{\omega}}$ and $\omega_i \equiv \frac{i}{N_{\omega}} \times \omega_{\text{max}}$. 
To get to the final form of the right-hand side of Eq. (\ref{eq:weight}), we made use of the fact that the 
measure $[d \rho]$ is defined as 
\begin{equation}
[d \rho] \equiv \displaystyle \prod_i \frac{d \rho_i}{\sqrt{\rho_i}}. \label{eq:measure}
\end{equation} 
\begin{figure}[t]
\centering
\includegraphics[width=8cm,clip]{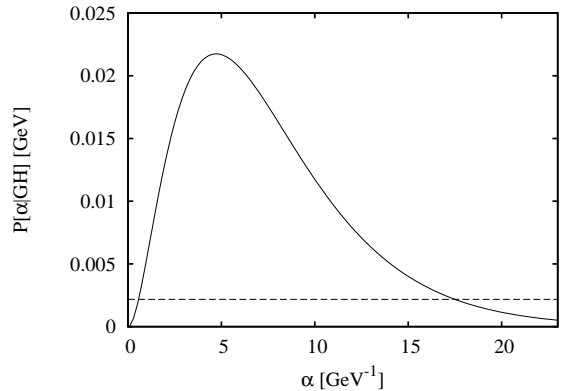}
\caption{Probability $P[\alpha|GH]$, as obtained in our analysis of mock data. To generate this 
function, the errors of Ref. \citen{Ioffe} and the default model of Eq. (\ref{eq:def.mod}) with 
parameters $\omega_0 = 2.0$ GeV and $\delta=0.1$ GeV have been used. The dashed line 
corresponds to the value of $0.1 \times P[\alpha_{\text{max}} |GH]$, from which the boundaries of the 
integration region, $\alpha_{\text{low}}$ and $\alpha_{\text{up}}$, are determined.}
\label{fig:alpha.prob}
\end{figure} 
In Eq. (\ref{eq:weight}), we still have to specify $P[\alpha |H]$, for which one uses either Laplace's rule with 
$P[\alpha |H] = \text{const.}$ or Jeffreys' rule with $P[\alpha |H] = \frac{1}{\alpha}$. We will employ Laplace's rule 
throughout this study, but have confirmed that the analysis using Jeffreys' rule gives essentially the same results, 
leading to a shift of the final $\rho$-meson peak of only 10 - 20 MeV.

To give an idea of the behavior of the probability $P[\alpha|GH]$, a typical example of this 
function is shown in Fig. \ref{fig:alpha.prob}. 
Its qualitative structure is the same for all the cases studied in this paper. 
To calculate the integral of Eq. (\ref{eq:average}), we first determine the maximum value of $P[\alpha |GH]$, 
which has a pronounced peak: 
$P[\alpha_{\text{max}} |GH]$. Then, we obtain the lower and upper boundaries of the integration region 
($\alpha_{\text{low}}$, $\alpha_{\text{up}}$) from the condition $P[\alpha |GH] > 0.1 \times P[\alpha_{\text{max}} |GH]$ 
and normalize $P[\alpha |GH]$ so that its integral within the integration region gives 1. After these 
preparations, the average of Eq. (\ref{eq:average}) is evaluated numerically. 

\subsubsection{Error estimation}  
As a last step of the MEM analysis, we have to estimate the error of the obtained result $\rho_{\mathrm{out}}(\omega)$. 
The error is calculated for averaged values of $\rho_{\mathrm{out}}(\omega)$ over a certain interval ($\omega_1$, $\omega_2$), 
as shown below.

The variance of $\rho(\omega)$ from its most probable form for fixed $\alpha$, 
$\delta \rho_{\alpha}(\omega) \equiv \rho(\omega) - \rho_{\alpha}(\omega)$, averaged over the 
interval ($\omega_1$, $\omega_2$), is defined as 
\begin{equation}
\begin{split}
\langle (\delta \rho_{\alpha})^2 \rangle_{\omega_1, \omega_2} &\equiv \frac{1}{(\omega_2 - \omega_1)^2}
\displaystyle \int [d \rho] \displaystyle \int_{\omega_1}^{\omega_2} d \omega d \omega^{\prime} 
\delta \rho_{\alpha}(\omega) \delta \rho_{\alpha}(\omega^{\prime}) P[\rho|GH] \\
&= - \frac{1}{(\omega_2 - \omega_1)^2} \displaystyle \int_{\omega_1}^{\omega_2} d \omega d \omega^{\prime} 
\biggl( \frac{\delta^2 Q}{\delta \rho(\omega) \delta \rho(\omega^{\prime})} \biggr)^{-1} \bigg|_{\rho=\rho_{\alpha}},
\end{split}
\label{eq:error} 
\end{equation}
where again the definition of Eq. (\ref{eq:measure}) and the Gaussian approximation 
for $P[\rho|GH]$ were used. Taking the root of this 
expression and averaging over $\alpha$ 
in the same way as was done for $\rho_{\alpha}(\omega)$, we obtain the final result 
of the error $\delta \rho_{\mathrm{out}}(\omega)$, averaged over the interval ($\omega_1$, $\omega_2$):
\begin{equation}
\langle \delta \rho_{\mathrm{out}} \rangle_{\omega_1, \omega_2} = \displaystyle \int_{\alpha_{\text{low}}}^{\alpha_{\text{up}}} d \alpha
\sqrt{\langle (\delta \rho_{\alpha})^2 \rangle_{\omega_1, \omega_2}} P[\alpha | G H].
\end{equation}
The interval ($\omega_1$, $\omega_2$) is usually taken to cover a peak or some other structure of interest. 
The formulas of this subsection will be used to generate the error bars of the various plots 
of $\rho_{\mathrm{out}}(\omega)$, as shown in the following sections.  

\section{Analysis using mock data}
The uncertainties that are involved in QCD sum rule calculations mainly originate from the 
ambiguities of the condensates and other parameters such as the strong coupling constant 
or the quark masses. These uncertainties usually lead to results with relative errors of about 
20\%. It is therefore not a trivial question if MEM can be used to analyze the QCD sum rule 
results, or if the involved uncertainties are too large to allow a sufficiently accurate 
application of the MEM procedure.
 
To investigate this question in detail, we carry out the MEM analysis using mock data and 
realistic errors. Furthermore, we will study the dependence of the results on various choices 
of the default model $m(\omega)$ and determine which one is the most suitable for our purposes. 
This analysis will also provide us with an estimate of the precision of the final 
results that can be achieved by this method and what kind of general structures of the spectral 
function can or cannot be reproduced by the MEM procedure. 

\subsection{Generating mock data and the corresponding errors}
Following Refs. \citen{Asakawa} and \citen{Shuryak}, we employ a relativistic Breit-Wigner peak and 
a smooth function describing the transition to the asymptotic value at high energies for 
our model spectral function of the $\rho$-meson channel:
\begin{equation}
\begin{split}
\rho_{\mathrm{in}}(\omega) =&  \frac{2F_{\rho}^2}{\pi} 
\frac{\Gamma_{\rho}m_{\rho}}{(\omega^2 - m^2_{\rho})^2 + \Gamma^2_{\rho} m^2_{\rho}} 
+\frac{1}{4\pi^2} \Bigl( 1 + \frac{\alpha_s}{\pi} \Bigl)\frac{1}{1+ e^{(\omega_0 -\omega) / \delta}}, \\
\Gamma_{\rho}(\omega) =& \frac{g^2_{\rho \pi \pi}}{48 \pi} m_{\rho} 
\Bigl( 1- \frac{4 m^2_{\pi}}{\omega^2} \Bigr)^{3/2} 
\theta(\omega - 2m_{\pi}).
\end{split}
\label{eq:spectralfunc}
\end{equation} 
The values used for the various parameters are
\begin{equation}
\begin{split}
m_{\rho} &= 0.77 \hspace{0.1cm} \mathrm{GeV,} \hspace{0.5cm} m_{\pi} = 0.14 \hspace{0.1cm} \mathrm{GeV,} \\
\omega_0 &= 1.3 \hspace{0.1cm} \mathrm{GeV,} \hspace{0.5cm} \delta = 0.2 \hspace{0.1cm} \mathrm{GeV,} \\
g_{\rho \pi \pi} &= 5.45, \hspace{0.5cm} \alpha_s = 0.5, \\
F_{\rho} &= \frac{m_{\rho}}{g_{\rho \pi \pi}} = 0.141 \hspace{0.1cm} \mathrm{GeV.}
\end{split}
\label{eq:spfpara}
\end{equation}
The spectral function of Eq. (\ref{eq:spectralfunc}) is then substituted into Eq. (\ref{eq:fsumr}) and the 
integration over $\omega$ is performed numerically to obtain the central values of the data points of 
$G_{\mathrm{mock}}(M)$. (In this section, we will use $G_{\mathrm{mock}}(M)$ instead of $G_{OPE}(M)$
to make it clear that we are analyzing mock data.) 

We now also have to put some errors $\sigma(M)$ to the function $G_{\mathrm{mock}}(M)$. To make the analysis as 
realistic as possible, we will use exactly the same errors as in the actual investigation of the OPE results. 
How these errors are obtained will be discussed later, in the section where the real OPE results are 
analyzed. We just mention here that when analyzing the OPE results, we will use three different parametrizations 
for the condensates and other parameters, namely, those given in Refs. \citen{Colangelo,Narison,Ioffe} 
(see Table \ref{parameters}). 
These parametrizations lead to 
different estimations of the errors, but for the mock data analysis of this section, these differences 
are not very important. Here, we will therefore mainly use the errors obtained from the parameters of 
Ref. \citen{Ioffe}. 
The resulting function $G_{\mathrm{mock}}(M)$ is given in Fig. \ref{fig:mockdata}, together with the range 
$G_{\mathrm{mock}}(M) \pm \sigma(M)$.

\subsubsection{Determination of the analyzed Borel mass region}
Next, we have to decide what range of $M$ to use for the analysis. For 
the lower boundary, we can use the usual convergence criterion of the OPE such that the contribution of the highest-dimensional 
operators is less than 10\% of the sum of all OPE terms. This is a reasonable choice, as the errors originating from the ranges 
of condensate values lead to uncertainties of up to 20\%, and it would therefore not make much sense to set up a more strict 
convergence criterion. 
For the parameters of Ref. \citen{Ioffe}, this criterion gives $M_{\mathrm{min}} = 0.77$ GeV. 

Considering the upper boundary of $M$, the situation is less clear. In the conventional QCD sum rule 
analysis, it is standard to use the pole dominance condition, which makes sure that the contribution of the continuum states does 
not become too large. As we do not resort to the ``pole + continuum" ansatz in the current approach, the pole dominance criterion 
does not have to be used and one can, in principle, choose any value for the upper boundary of $M$. Nevertheless, 
because we are mainly interested in the lowest resonance peak, we will use a similar pole dominance criterion as in the 
traditional QCD sum rules.
\begin{figure}[t]
\centering
\includegraphics[width=8cm,clip]{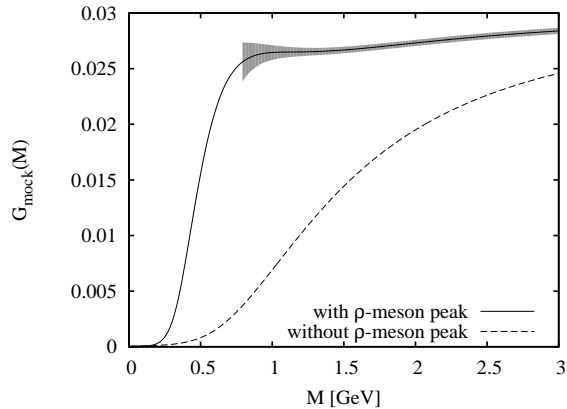}
\caption{Central values of the mock data obtained by numerically integrating Eq. (\ref{eq:fsumr}) with 
Eq. (\ref{eq:spectralfunc}). The errors of the data, extracted from the parameters of Ref. \citen{Ioffe}, are indicated 
by the gray region. The lower boundary of the shown errors corresponds to $M_{\mathrm{min}}$, below which 
the OPE does not converge. For comparison, the integral of Eq. (\ref{eq:fsumr}) is also shown for the case when only 
the continuum part, the second term of Eq. (\ref{eq:spectralfunc}), is taken for the spectral function.}
\label{fig:mockdata}
\end{figure}
By examining the 
mock data in Fig. \ref{fig:mockdata}, one sees that while the resonance pole contributes most strongly to the data around $M \sim 1$ GeV, the 
contributions from the continuum states grow with increasing $M$ and finally start to dominate the data for values that 
are larger than $1.5$ GeV. We will therefore use $M_{\mathrm{max}}=1.5$ GeV as the upper boundary of $M$ for the rest of the 
present paper. 
The dependence of the final results 
on this choice is small, as will be shown later in Fig. \ref{fig:Mmax}.

Finally, the values of the $N_M$ data points of $G_{\mathrm{mock}}(M)$ between $M_{\mathrm{min}}$ 
and $M_{\mathrm{max}}$ are randomly generated, using Gaussian 
distributions with standard deviations $\sigma(M)$, 
cenered at the values obtained from the integration of 
Eq. (\ref{eq:fsumr}). 
The ranges of values of $G_{\mathrm{mock}}(M)$ are indicated by 
the gray region in Fig. \ref{fig:mockdata}. 
We take 100 data points for functions of the Borel mass $M$ ($N_M=100$) and 
have checked that the results of the analysis do not change when this 
value is altered. 
For functions of the energy $\omega$, 600 data points are taken ($N_{\omega}=600$).

\subsection{Choice of an appropriate default model}
It is important to understand the meaning of the default model $m(\omega)$ in 
the present calculation. It is in fact used to fix the value of the spectral function at high and 
low energies, because 
the function $G_{OPE}(M)$ contains only little information on these regions. This can be understood 
firstly by considering 
the property of the kernel of Eq. (\ref{eq:kernel}), which is zero at $\omega = 0$. $G_{OPE}(M)$ 
therefore contains no information on $\rho_{\mathrm{in}}(\omega=0)$ 
and the corresponding 
result of the MEM analysis $\rho_{\mathrm{out}}(\omega=0)$ will thus always approach 
the default model, as $G_{OPE}(M)$ does not constrain its value. 
Secondly, we use $G_{OPE}(M)$ only in a limited 
range of $M$, because the operator product expansion diverges for small $M$ and the region of very high 
$G_{OPE}(M)$ is dominated by the high-energy continuum states, which we are not interested in. 
The region of the spectral function, which contributes most strongly to $G_{OPE}(M)$ between $M_{\mathrm{min}}$ and 
$M_{\mathrm{max}}$, lies roughly in the range 
between $\omega_{\mathrm{min}}$ ($\simeq M_{\mathrm{min}}$) and 
$\omega_{\mathrm{max}}$ ($\simeq M_{\mathrm{max}}$), as can be for instance 
inferred from Fig. \ref{fig:mockdata}.
$\rho_{\mathrm{out}}(\omega)$ will then 
approach the default model quite quickly outside of this region, because there is no strong constraint 
from $G_{OPE}(M)$. 
This implies that the values of $\rho_{\mathrm{out}}(\omega)$ at the boundaries 
are fixed by the choice of the default model and one should therefore consider these boundary conditions as 
inputs of the present analysis. Once these limiting values of $\rho(\omega)$ are chosen, 
the MEM procedure then extracts the most probable spectral function $\rho_{\mathrm{out}}(\omega)$ given $G_{OPE}(M)$ 
and the boundary conditions supplied by $m(\omega)$.

\begin{figure}[t]
\centering
\includegraphics[width=8cm,clip]{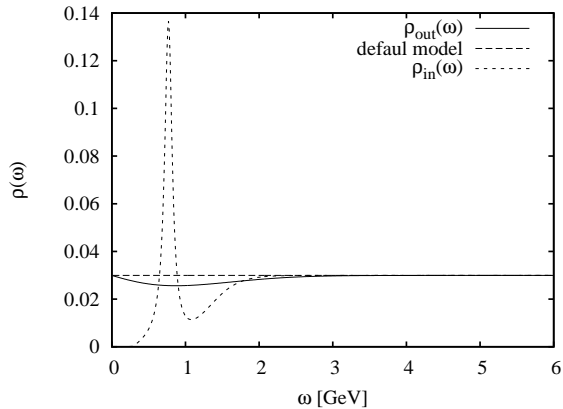}
\caption{Outcome of the MEM analysis using a constant default model with its value fixed to 
the perturbative result. $\rho_{\mathrm{in}}(\omega)$ is the function that was used to produce the 
mock data, given in Eq. (\ref{eq:spectralfunc}), and $\rho_{\mathrm{out}}(\omega)$ shows the 
result of the MEM procedure.}
\label{fig:const}
\end{figure}

To illustrate the importance of choosing appropriate boundary conditions, 
we show the results of the MEM analysis for a constant default model, 
with a value fixed to the perturbative result at high energy. 
Here, the boundary condition for the low energy is not correctly 
chosen, because the spectral function is expected to vanish at very low energy. 
The result is given in Fig. \ref{fig:const} and clearly shows that 
$\rho_{\mathrm{out}}(\omega)$ does not reproduce $\rho_{\mathrm{in}}(\omega)$. 

This is in contrast to the corresponding behavior in lattice QCD, where it suffices
to take a constant value of the spectral function, chosen to be consistent
with the high-energy behavior of the spectral function to obtain correct 
results. The reason for this difference is mainly that the OPE in QCD sum rules 
is not sensitive to 
the low-energy part of the spectral function, 
owing to the properties of the 
kernel and our limitations in the applicability of the OPE. 
Most importantly, the information that there is essentially
no strength in the spectral function below the rho-meson peak, is stored in the
region of the Borel mass $M$ around and below $M=0.5$ GeV. However, as the OPE does not
converge in this region, it is not available for our analysis and we therefore
have to use the default model to adjust the spectral function to the correct
behavior. 
On the other hand, in lattice QCD, it is possible to calculate the
correlator reliably at sufficiently high euclidean time, where the effective
mass plot reaches a plateau and thus only the ground state contributes. Therefore,
it seems that from lattice QCD one can gain sufficient information on the 
low-energy part of the spectral function, and one does not need to adjust the
default model to obtain physically reasonable results.

For the present analysis, we introduce 
the following functional form, to smoothly connect low- and high-energy parts of the default model, 
\begin{equation}
m(\omega) = \frac{1}{4\pi^2}\Bigl( 1 + \frac{\alpha_s}{\pi} \Bigl)\frac{1}{1+ e^{(\omega_0 -\omega) / \delta}}, \label{eq:def.mod}
\end{equation}
which is close to 0 at low energy and approaches the perturbative value $1 + \frac{\alpha_s}{\pi}$ at high 
energy, changing most significantly in the region between $\omega_0 - \delta$ and $\omega_0 + \delta$. 
This function can be considered to be the counterpart of the ``continuum" in the ``pole + continuum" 
assumption of Eq. (\ref{eq:pole}), 
where $\delta$ is essentially taken to be 0 and the threshold parameter $s_{\mathrm{th}}$ 
corresponds to $\omega_0$. 
Equation (\ref{eq:def.mod}) nevertheless enters into the calculation in a very different way than the second 
term of Eq. (\ref{eq:pole}) in the conventional sum rules, and one should therefore not regard 
these two approaches to the continuum as completely equivalent. 

\begin{figure}[t]
\centering
\includegraphics[width=14cm,clip]{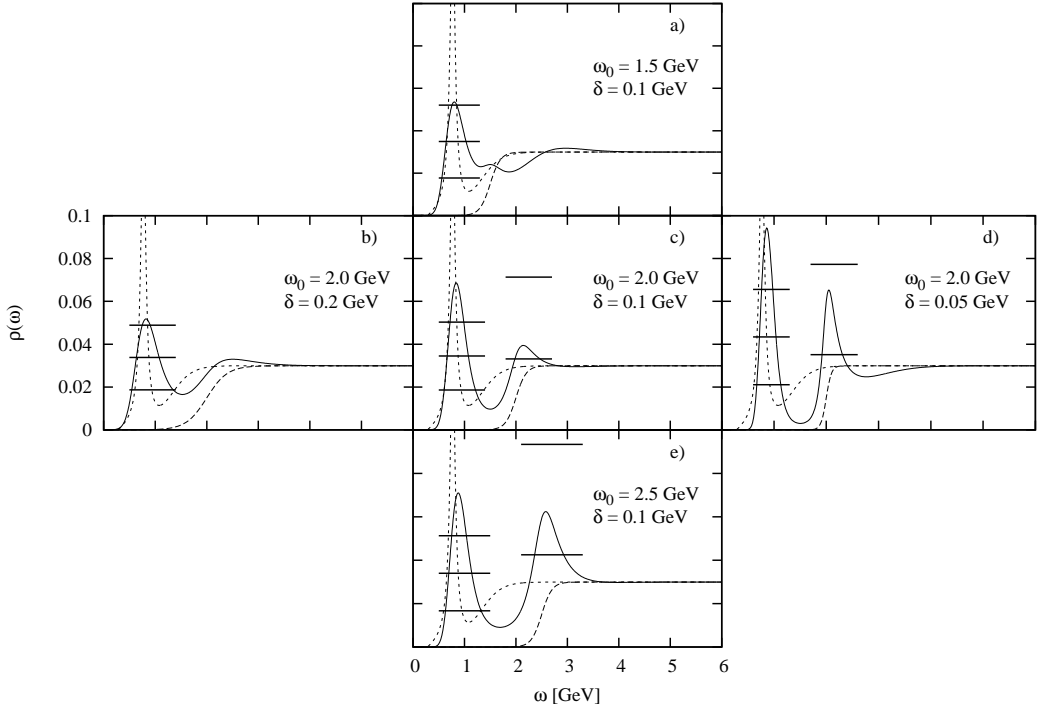} 
\caption{Results of the MEM investigation of mock data with various default models. As in Fig. \ref{fig:const}, 
the solid lines stand for the output of the analysis $\rho_{\mathrm{out}}(\omega)$, the long-dashed lines for 
the default model with the parameters shown in the figure, and the short-dashed lines for the input function 
$\rho_{\mathrm{in}}(\omega)$ of Eq. (\ref{eq:spectralfunc}). The horizontal bars show the values of the spectral 
function, averaged over the peaks $\langle \rho_{\mathrm{out}} \rangle_{\omega_1 ,\omega_2}$ and the 
corresponding ranges 
$\langle \rho_{\mathrm{out}} \rangle_{\omega_1 ,\omega_2} \pm \langle \delta \rho_{\mathrm{out}} \rangle_{\omega_1 ,\omega_2}$. 
Their extent shows the averaged interval $(\omega_1, \omega_2)$. 
For figures c), d) and e), the lower error bars 
of the second peak are not shown because they lie below $\rho(\omega) = 0$.}
\label{fig:mock.default}
\end{figure}

We have tested the MEM analysis of the mock data for 
several values of $\omega_0$ and $\delta$ and the results are shown in Fig. \ref{fig:mock.default}. 
One sees that in the cases of d) and e), the sharply rising default model induces an 
artificial peak in the region of $\omega_0$. Even though these peaks are statistically not 
significant, they may lead to erroneous conclusions. 
We will therefore adopt only 
default models, for which only small artificial structures appear, 
such as in the case of the figures a), b) and c). Comparing these 
three figures, it is observed   
that the error of the spectral function relative to the height of the $\rho$-meson peak 
is smallest for the parameters of c). We therefore adopt this default model with 
$\omega_0=2.0$ GeV and $\delta=0.1$ GeV in the following investigations.  

It is worth considering the results of Fig. \ref{fig:mock.default} also from the viewpoint of the dependence of the peak position 
on the default model. It is observed that even though the height of the $\rho$-meson peak varies quite strongly 
depending on which default model is chosen, its position only varies in a range of $\pm 40$ MeV, 
which shows that the present evaluation of the lowest pole position does not strongly depend on the detailed values of 
$\omega_0$ and $\delta$. 
This behavior should be compared with the results of the usual sum rules, where the 
dependence of the pole mass on the 
threshold parameter $s_{\mathrm{th}}$ is stronger. 
On the other hand, we have to mention that $\omega_0$ should not be chosen to have a value much below $\omega_0 = 1.5$ GeV, 
because, in this region, the artificial structures discussed above start to interfere with the $\rho$-meson peak. Moreover, 
if the default model approaches the limit shown in Fig. \ref{fig:const}, where $m(\omega)$ is just a constant fixed to 
the asymptotic value at high energy, the $\rho$-meson peak will gradually disappear. 

\subsection{Investigation of the stability of the obtained spectral function}
\begin{figure}[b]
\centering
\includegraphics[width=8cm,clip]{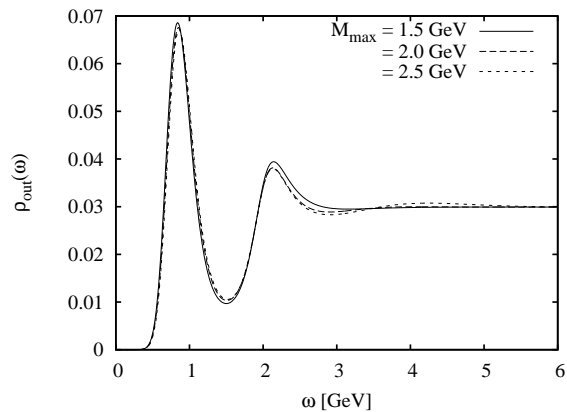}
\caption{MEM results for three different values of the upper boundary of the Borel mass $M_{\mathrm{max}}$. 
For the default model, the same version as in part c) of Fig. \ref{fig:mock.default} was used. Thus, 
the solid line of this figure is the same as the one of part c) in Fig. \ref{fig:mock.default}.}
\label{fig:Mmax}
\end{figure}

Next, we will briefly discuss the dependence of our results on the upper boundary of the employed Borel mass region 
$M_{\mathrm{max}}$. 
In Fig. \ref{fig:Mmax}, we show the results for the values $M_{\mathrm{max}}=1.5$ GeV, $2.0$ GeV and $2.5$ GeV. 
Here, the default model with parameters $\omega_0=2.0$ GeV and $\delta=0.1$ GeV was used.
The spectral functions of these three cases almost coincide and have the same 
qualitative features. Quantitatively, the peak position of the $\rho$-meson is shifted only 20 MeV when 
$M_{\mathrm{max}}$ is raised from $1.5$ GeV to $2.5$ GeV.

\begin{figure}
\centering
\includegraphics[width=8cm,clip]{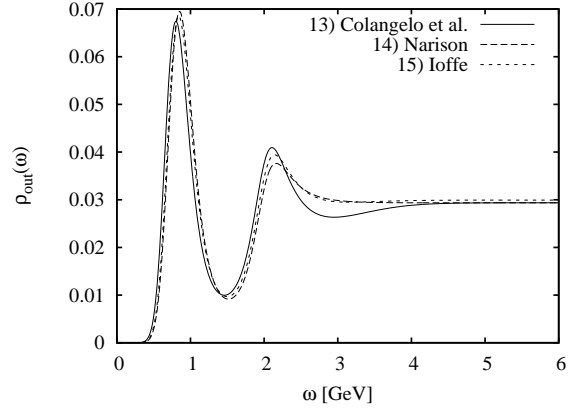}
\caption{Findings of the MEM analysis with different error estimations and lower 
boundaries of the Borel mass $M_{\mathrm{min}}$. 
For the default model, the same version as in part c) of Fig. \ref{fig:mock.default} was used.
The solid line uses parameters of Ref. \citen{Colangelo}, 
the long-dashed lines those of Ref. \citen{Narison} and the short-dashed line those of Ref. \citen{Ioffe}. 
Note that we use for this plot the same mock data for all three cases and vary only 
the errors and $M_{\mathrm{min}}$.}
\label{fig:errordependence}
\end{figure}

Let us also check how the results of the analysis are affected by 
a different choice of parameters, leading to altered magnitudes of error and 
also differing lower boundaries of the Borel mass $M_{\mathrm{min}}$. 
Of the three parameter sets used in this study, 
given in Table \ref{parameters}, the errors of Ref. 
\citen{Colangelo} are rather small, while the errors of Refs. \citen{Narison} 
and \citen{Ioffe} are larger 
and have about the same overall magnitude. 
Moreover, $M_{\mathrm{min}}$, which is determined from the OPE convergence criterion 
mentioned earlier, takes values $M_{\mathrm{min}}=0.72$ GeV for Ref. \citen{Colangelo}, 
$0.83$ GeV for Ref. \citen{Narison} and $0.77$ GeV for Ref. \citen{Ioffe}.
To understand how these parameters affect the MEM analysis, 
the results of the calculation using the same central values for the mock data, but 
different errors and $M_{\mathrm{min}}$, are shown in Fig. \ref{fig:errordependence}. 
It is observed that the spectral functions are very similar and depend only 
weakly on the choice of errors and $M_{\mathrm{min}}$.

\begin{figure}[t]
\centering
\includegraphics[width=8cm,clip]{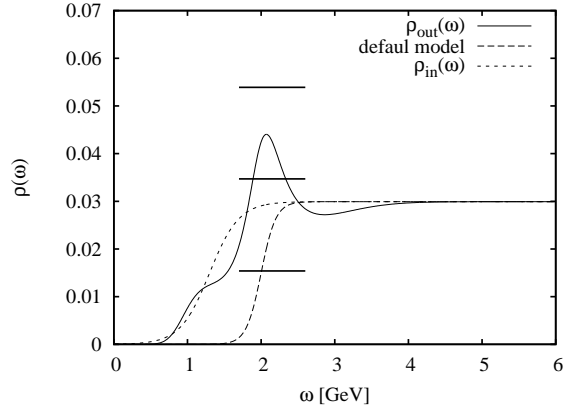}
\caption{Outcome of the MEM analysis using mock data without the $\rho$-meson peak. 
$\rho_{\mathrm{in}}(\omega)$ is the function that was used to produce the 
mock data and $\rho_{\mathrm{out}}(\omega)$ shows the 
result of the MEM procedure. 
For the default model, the same version as in part c) of Fig. \ref{fig:mock.default} was used.}
\label{fig:nopeak}
\end{figure}

Finally, it is important to confirm whether the lowest peak that we have observed 
in all the results so far really originates from the $\rho$-meson peak 
of the input spectral function. In other words, we have to verify that the 
lowest peak obtained from the MEM analysis really disappears when the 
$\rho$-meson peak is removed from the input spectral function. The result for 
this case is given in Fig. \ref{fig:nopeak}.    
We see that while we get the same (nonsignificant) peak around $2.0$ GeV as before, which is induced by 
the sharply rising default model in this region, the lower peak has completely 
disappeared. This confirms that the lower peak is directly related to the $\rho$-meson peak 
and is not generated by any other effects of the calculation.

\subsection{Estimation of the precision of the final results}
To obtain an estimate of the precision of the current approach, 
let us now turn to the numerical results of our analysis of 
mock data. We regard plot c) of Fig. \ref{fig:mock.default} as our main result, the central value of the 
peak being $m_{\rho, \mathrm{out}}=0.84$ GeV. The shift from the true value of $m_{\rho, \mathrm{in}}=0.77$ GeV is 
caused by the errors of the involved parameters and by the fact that we cannot use all the data points 
of $G_{\mathrm{mock}}(M)$, but only the ones for which we have some confidence that the OPE converges.
 
Furthermore, as discussed above, there are some additional uncertainties of $\pm 40$ MeV coming from the choice 
of the default model and $\pm 20$ MeV from the value of $M_{\mathrm{max}}$. 
The overall error is then obtained by taking the root
of the sum of all squared errors and rounding it up. 
This gives 
\begin{equation}
\Delta m_{\rho} \simeq  \pm 90 \, \mathrm{MeV},
\end{equation}  
which is quite large but seems to be realistic when one considers the large errors of the condensates 
that are involved in the calculation. For the other parametrizations, we get similar results, concretely, 
$\Delta m_{\rho} \simeq  \pm 60$ MeV for Ref. \citen{Colangelo} and $\Delta m_{\rho} \simeq  \pm 100$ MeV for 
Ref. \citen{Narison}.
 
\begin{figure}[t]
\centering
\includegraphics[width=8cm,clip]{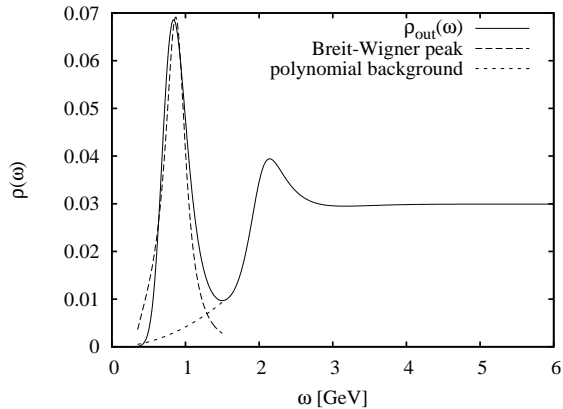}
\caption{An example of a result of the fitting procedure described in the text. For the spectral function 
$\rho_{\mathrm{out}}(\omega)$, we use here the same as in part c) of Fig. \ref{fig:mock.default}. For the 
peak, a relativistic Breit-Wigner form is employed, while the background is parametrized by a second-order 
polynomial with positive valued coefficients.}
\label{fig:fitresult}
\end{figure}

Having the spectral function at our disposal, it becomes possible to extract the coupling strength of the interpolating 
field to the $\rho$-meson state, the quantity corresponding to $F_{\rho}$ in Eq. (\ref{eq:spectralfunc}). We obtain 
this coupling strength by fitting the spectral function in the region of the $\rho$-meson resonance with a 
relativistic Breit-Wigner peak of the same functional form as 
the first term of Eq. (\ref{eq:spectralfunc}) 
plus a second-order polynomial to describe the continuum background. 
In order that the background does not become negative and does not contribute to the peak, we 
have restricted the coefficients of the polynomial to positive values. 
An example of the resulting curves is given in Fig. \ref{fig:fitresult}. 
For the case of plot c) in Fig. \ref{fig:mock.default}, we get a value of $F_{\rho, \mathrm{out}}=0.178$ GeV, which 
is somewhat larger than the true value of $F_{\rho, \mathrm{in}}=0.141$ GeV. It is not a surprise that the precision 
of this quantity is worse than that of the peak position, because the shape of the peak is deformed quite 
strongly owing to the MEM procedure. 
As can be suspected when looking at Fig. \ref{fig:mock.default},
there is also a fairly large uncertainty coming from the choice of the default model, which is about $\pm 0.031$ GeV. 
On the other hand, we found that the dependence on the boundaries of the fitting region and on $M_{\mathrm{max}}$ 
is very small, being $\pm 0.003$ GeV and $\pm 0.001$ GeV, respectively. Altogether, 
this gives the following error for $F_{\rho}$:
\begin{equation}
\Delta F_{\rho} \simeq  \pm 0.049 \, \mathrm{GeV}.
\end{equation}
A similar analysis for the parameters of Ref. \citen{Colangelo} gives $\Delta F_{\rho} \simeq  \pm 0.038$ GeV and 
$\Delta F_{\rho} \simeq  \pm 0.049$ GeV for Ref. \citen{Narison}.
\begin{figure}[t]
\centering
\includegraphics[width=8cm,clip]{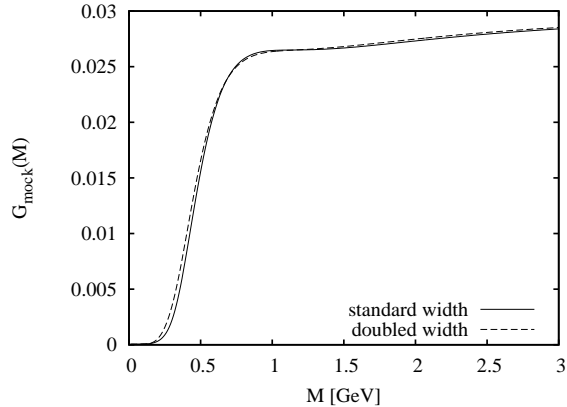}
\caption{Two versions of $G_{\mathrm{mock}}(M)$, the solid line showing the one with the standard value for the 
width as given in Eqs. (\ref{eq:spectralfunc}) and (\ref{eq:spfpara}), and the dashed line giving a version 
where $\Gamma_{\rho}(\omega)$ of Eq. (\ref{eq:spectralfunc}) is doubled and $F_{\rho}=0.149$ GeV is used.}
\label{fig:width}
\end{figure}

As one further point, it is important to investigate if and how the precision of the MEM analysis improves once 
the OPE data will be available with better precision. To answer this question, 
we have repeated the analysis using an error with a smaller magnitude and have found that, concerning the pole position, 
the reproducibility indeed improves with a smaller error. For instance, using the errors obtained from the Ioffe parameters 
of Ref. \citen{Ioffe}, 
we get $m_{\rho,\mathrm{out}}=0.84$ GeV, but when we reduce the errors by hand to 20\% of their real value, the result 
shifts to $m_{\rho,\mathrm{out}}=0.78$ GeV, which almost coincides with the correct value of $m_{\rho,\mathrm{in}}=0.77$ GeV. 
To a lesser degree, the same is true for the residue. Its value changes from $F_{\rho,\mathrm{out}}=0.178$ GeV 
to $F_{\rho,\mathrm{out}}=$ 0.167 GeV with reduced error values, compared with the input value of 
$F_{\rho,\mathrm{in}}=0.141$ GeV, which is also an improvement. 
On the other hand, we could not observe a significantly better reproducibility of the width with the reduced error values.

These results show that the MEM analysis of QCD sum rules has the potential to become more accurate in obtaining the 
position and the residue of the $\rho$-meson pole, once the values of the condensates are known with better precision. 
At the same time, it has to be noted that an accurate determination of the width seems to be difficult to achieve 
in the current approach even with smaller errors. The reason for this is discussed in the following section.

\subsection{Why it is difficult to accurately determine the width of the $\rho$-meson}
We have so far focused our discussion on the reproduction of the parameters $m_{\rho}$ and $F_{\rho}$ of the spectral 
function and have shown that, by the MEM analysis, they could be reproduced with a precision of approximately 20\%. 
Considering now the width $\Gamma_{\rho}$, the situation turns out to be quite different, as can be observed for 
instance in Fig. \ref{fig:mock.default}. We see there that the values of the width come out about twice as large as 
in the input spectral function, Eq. (\ref{eq:spectralfunc}). 

The reason for this difficulty of reproducing the width lies in the small dependence of $G_{\rho}(M)$ on 
$\Gamma_{\rho}$. This is illustrated in Fig. \ref{fig:width}. 
We see there that the curve obtained from our model spectral function of Eqs. (\ref{eq:spectralfunc}) and 
(\ref{eq:spfpara}) and the one generated from a spectral function for which the width of the $\rho$-meson peak 
has been doubled and a slightly larger value for $F_{\rho}$ has been used ($F_{\rho}=0.149$ GeV compared with 
the standard value of $0.141$ GeV), almost 
coincide. This means that with the precision available for our QCD sum rule analysis, it is practically 
impossible to distinguish between these two cases. Examining the curves of Fig. \ref{fig:width} a bit more 
carefully, it is found that the most prominent difference between them lies in the region 
of $0.5$ GeV or below. However, this region cannot be accessed by the current calculation, as the OPE 
does not converge well for such small values of $M$. Furthermore, even if we could have calculated 
the OPE to higher orders and would thus have some knowledge about $G_{OPE}(M)$ in the region below $0.5$ GeV, 
this would most likely not help much, as the uncertainty here will be large owing to the large number 
of unknown condensates that will appear at higher orders of the OPE. 
We therefore have to conclude that it is 
not possible to say anything meaningful about the width of the $\rho$-meson peak at the current 
stage. To predict the width, the OPE 
has to be computed to higher orders and the various condensates have to be known with much 
better precision than they are today.   

\section{Analysis using the OPE results}
\subsection{The $\rho$-meson sum rule}
Carrying out the OPE and applying the Borel transformation, we obtain $G_{OPE}(M)$, the left-hand 
side of Eq. (\ref{eq:fsumr}). In the case of the vector meson channel, we use the operator
\begin{equation}
j_{\mu} = \frac{1}{2}(\bar{u}\gamma_{\mu}u - \bar{d}\gamma_{\mu}d),  \label{eq:operator}
\end{equation}
which stands for $J$ in Eq. (\ref{eq:cor1}) and take the terms proportional to the structure 
$q_{\mu}q_{\nu} - q^2 g_{\mu\nu}$. We then arrive at the following expression for $G_{OPE}(M)$, 
where the OPE has been calculated up to dimension 6:
\begin{equation}
\begin{split}
G_{OPE}(M) =& \frac{1}{4 \pi^2} \Big( 1 + \eta(\alpha_s) \Big)  
+\Big( 2m \langle \bar{q} q \rangle  +  \frac{1}{12} \big \langle \frac{\alpha_s}{\pi} G^2 \big \rangle \Big) \frac{1}{M^4} \\
&- \frac{112 \pi}{81} \alpha_s \kappa \langle \bar{q} q \rangle^2 \frac{1}{M^6} + \dots, \\
\eta(\alpha_s) =&  \frac{\alpha_s}{\pi} + 0.154 \alpha^2_s - 0.372 \alpha^3_s + \dots.
\end{split}
\label{eq:operesult}
\end{equation}
Here, $\alpha_s$ is the usual strong coupling constant, given as $\frac{g^2}{4\pi}$, 
$m$ stands for the (averaged) quark mass of the \textit{u}- and \textit{d}-quarks, 
and $\langle \bar{q} q \rangle$ is the corresponding 
quark condensate. Meanwhile, the gluon condensate $\big \langle \frac{\alpha_s}{\pi} G^2 \big \rangle$ is an abbreviated expression for 
$\big \langle \frac{\alpha_s}{\pi} G^a_{\mu\nu} G^{a\mu\nu} \big \rangle$ and $\kappa$ parametrizes the breaking of the vacuum 
saturation approximation, which has been used to obtain the above result for $\kappa = 1$. 
\begin{table}[b]
\begin{center}
\caption{Values of the parameters used in the calculation. These have been adjusted to the renormalization scale of 1 GeV.} 
\label{parameters}
\begin{tabular}{lccc} \hline
 & Colangelo and Khodjamirian \cite{Colangelo} & \hspace*{1cm} Narison \cite{Narison} \hspace*{1cm}& Ioffe \cite{Ioffe} \\ \hline 
$\langle \bar{q} q \rangle$ & $-(0.24 \pm 0.01)^3\,\,\mathrm{GeV}^3$ 
& $-(0.238 \pm 0.014)^3\,\,\mathrm{GeV}^3$ & $-(0.248 \pm 0.013)^3\,\,\mathrm{GeV}^3$  \\
$\big \langle \frac{\alpha_s}{\pi} G^2 \big \rangle$ 
& $0.012 \pm 0.0036\,\,\mathrm{GeV}^4$ & $0.0226 \pm 0.0029\,\,\mathrm{GeV}^4$ & $0.009 \pm 0.007\,\,\mathrm{GeV}^4$  \\ 
$\kappa$ & 1 & $2.5 \pm 0.5$ & $1.0 \pm 0.1$  \\
$\alpha_s$ & $0.5$ & $0.50 \pm 0.07$ & $0.57 \pm 0.08$  \\ \hline
\end{tabular}
\end{center}
\end{table} 

A few comments about this result are in order here. For the perturbative term, which is known up to the third order in 
$\alpha_s$, we have taken the number of flavors to be $N_f = 4$. Note that only the second and third terms of $\eta(\alpha_s)$ 
depend (weakly) on $N_f$ \cite{Surguladze} and that the final results of the analysis are thus not affected by this choice. 
We have not considered the running of $\alpha_s$ in deriving Eq. (\ref{eq:operesult}) for simplicity. 
If the running is taken into account, the coefficient of the third term of $\eta(\alpha_s)$ changes due 
to the Borel transformation, as was shown in Ref. \citen{Shifman2}. 
Nevertheless, this again leads only to a minor 
change in the whole expression of Eq. (\ref{eq:operesult}) and does not alter any of the results 
shown in the following. 
Considering the terms proportional to $1/M^4$, the first-order corrections of the 
Wilson coefficients are in fact known \cite{Surguladze}, but we have 
not included them here as the values of the condensates themselves have large uncertainties, and 
it is therefore not necessary to determine the corresponding coefficients with such a 
high precision. It is nonetheless important to note that these corrections are small 
(namely of the order of 10\% or smaller, compared with the leading terms) and thus 
do not introduce any drastic changes into the sum rules. 

\subsubsection{Values of the parameters and their uncertainties}
There are various estimates of the values of the condensates and their ranges. We will employ the ones given in three recent 
publications: \citen{Colangelo,Narison,Ioffe}. The explicit values are given in Table \ref{parameters}, where they have been adjusted 
to the renormalization scale of 1 GeV. 
For the central value of $m \langle \bar{q} q \rangle$, we make use of the Gell-Mann-Oakes-Renner relation, which gives 
$m \langle \bar{q} q \rangle = -\frac{1}{2} m^2_{\pi} f^2_{\pi}$ and take the experimental values of 
$m_{\pi}$ and $f_{\pi}$ for all three cases, leading to 
\begin{equation}
m \langle \bar{q} q \rangle = -8.5 \times 10^{-5}\,\mathrm{GeV}^4.
\end{equation} 
Note that due 
to its smallness, the term containing $m \langle \bar{q} q \rangle$ does not play an important role 
in the sum rules of Eq. (\ref{eq:operesult}). 
The values of Table \ref{parameters} agree well for $\langle \bar{q} q \rangle$ and $\alpha_s$, while they differ considerably for 
$\big \langle \frac{\alpha_s}{\pi} G^2 \big \rangle$ and $\kappa$. Namely, Ref. \citen{Narison} employs values 
for $\big \langle \frac{\alpha_s}{\pi} G^2 \big \rangle$ and $\kappa$ that are about two 
times larger than those of Refs. \citen{Colangelo} and \citen{Ioffe}. Considering the error estimates of the parameters, 
Ref. \citen{Colangelo} uses altogether the smallest errors as the breaking of the vacuum saturation approximation is not considered and 
only a fixed value for $\alpha_s$ is employed. Comparing the results obtained from these three parameter sets will provide us 
with an estimate of the order of the systematic error inherent in the current calculation.  

\subsubsection{Determination of the errors of $G_{OPE}(M)$}
\begin{figure}[b]
\centering
\includegraphics[width=8cm,clip]{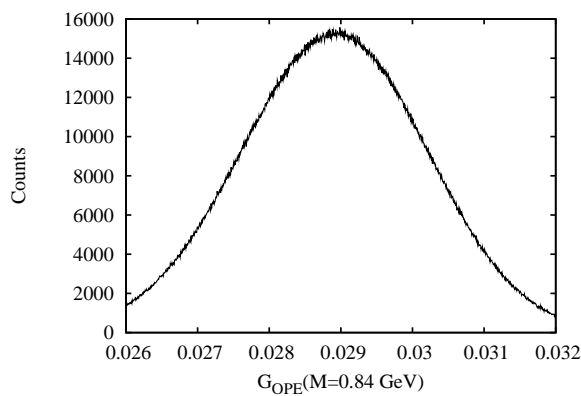}
\caption{An example of the distributed values of $G_{OPE}(M)$ for $M=0.84$ GeV. Here, the 
parameters of Ref. \citen{Ioffe} have been used.}
\label{fig:errorestimation}
\end{figure}

As can be inferred from Eq. (\ref{eq:operesult}) and Table \ref{parameters}, the uncertainty of $G_{OPE}(M)$ will vary as a 
function of $M$ and will be larger for small values of $M$ because the contributions of the higher-order terms with 
large uncertainties of the condensates become more important in that region. 
To accurately estimate this error, we follow Ref. \citen{Leinweber} and numerically generate Gaussianly distributed values for the condensates, 
and then examine the distribution of the resulting values of $G_{OPE}(M)$. For illustration, we show the values of 
$G_{OPE}(M=0.84 \, \mathrm{GeV})$ in Fig. \ref{fig:errorestimation}, where the parameter estimates of Ref. \citen{Ioffe} have been used. 
$\sigma(M)$, 
the error of $G_{OPE}(M)$, can be easily obtained from the formula of the standard deviation of a given set. 

As for the analysis of the mock data, the data points of $G_{OPE}(M)$ are randomly generated, using Gaussian 
distributions with standard deviations $\sigma(M)$, centered at $G_{OPE}(M)$ of Eq. (\ref{eq:operesult}). 
We here again take $N_M=100$ and $N_{\omega}=600$. 
$M_{\mathrm{min}}$ is determined from the 10\% convergence criterion and $M_{\mathrm{max}}$ is 
fixed at $1.5$ GeV.

\subsection{Results of the MEM analysis}
\begin{figure}[b]
\centering
\begin{tabular}{cc}
\includegraphics[width=6.7cm,clip]{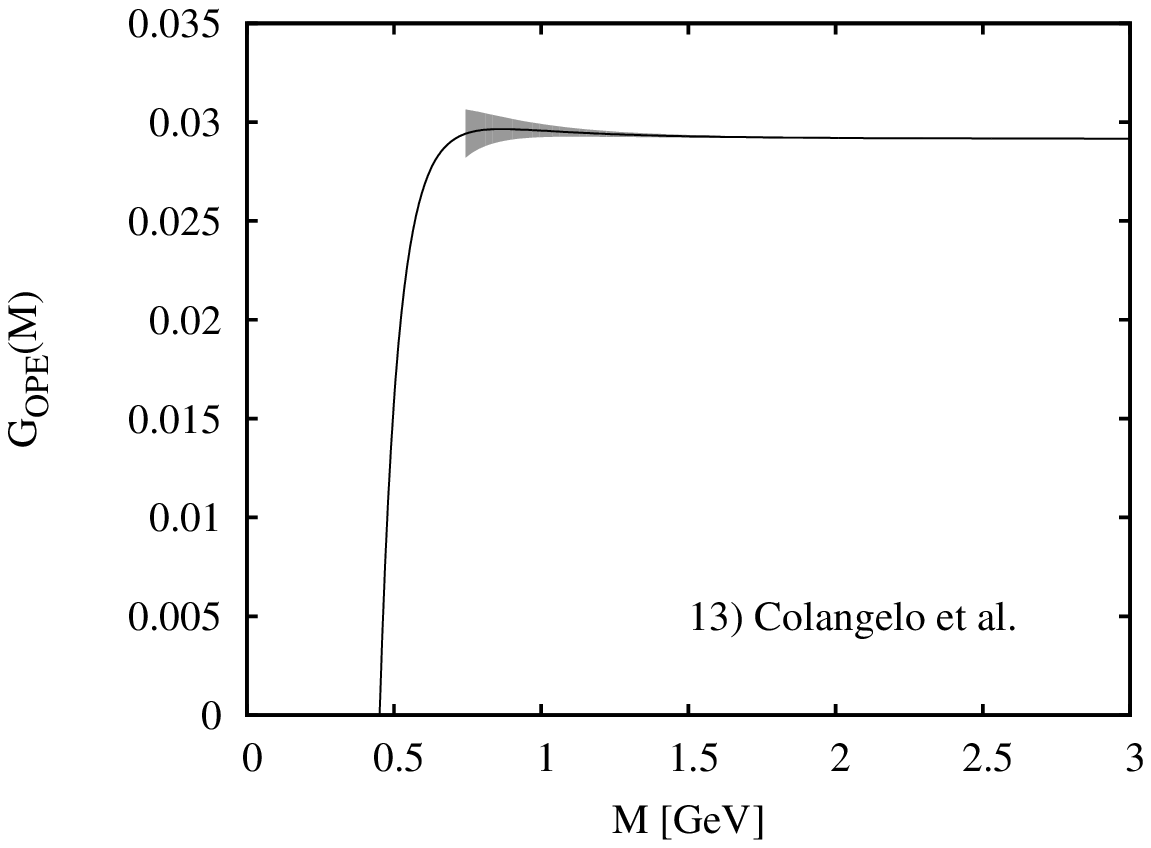} &  \includegraphics[width=6.7cm,clip]{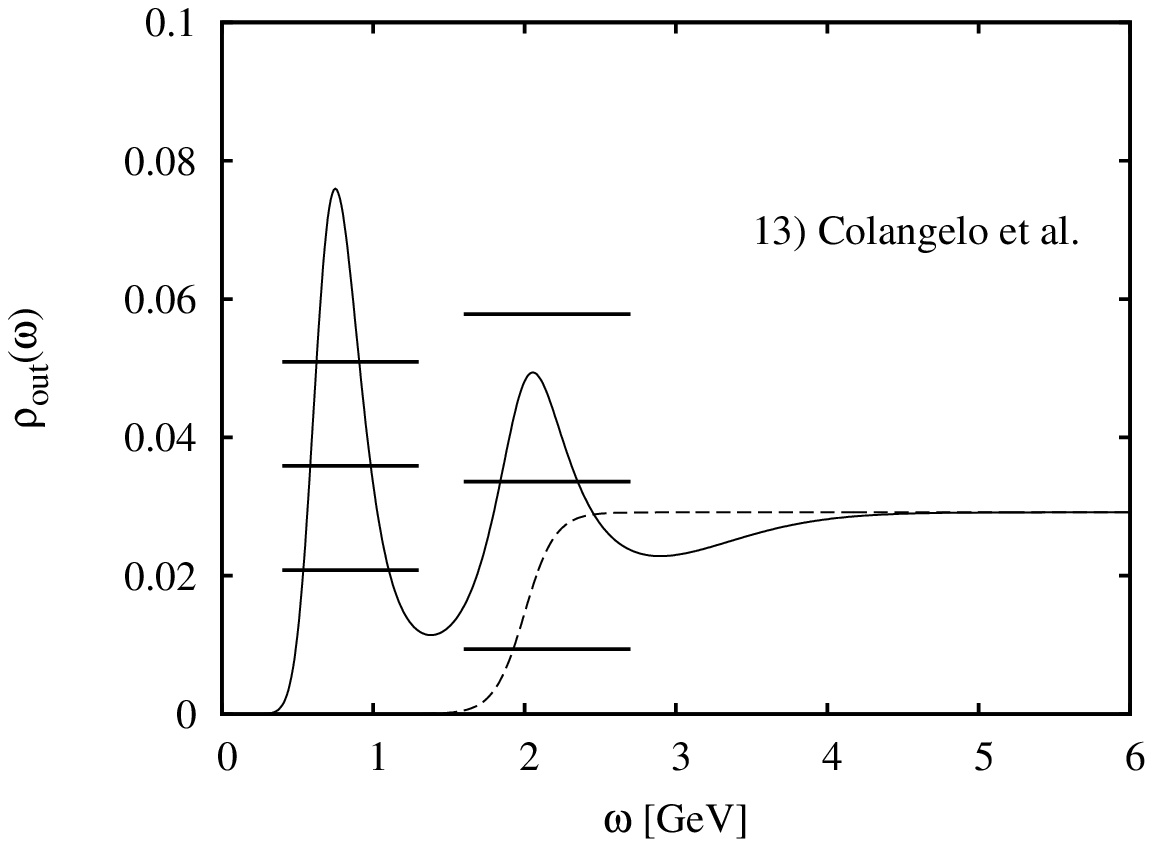} \\
\includegraphics[width=6.7cm,clip]{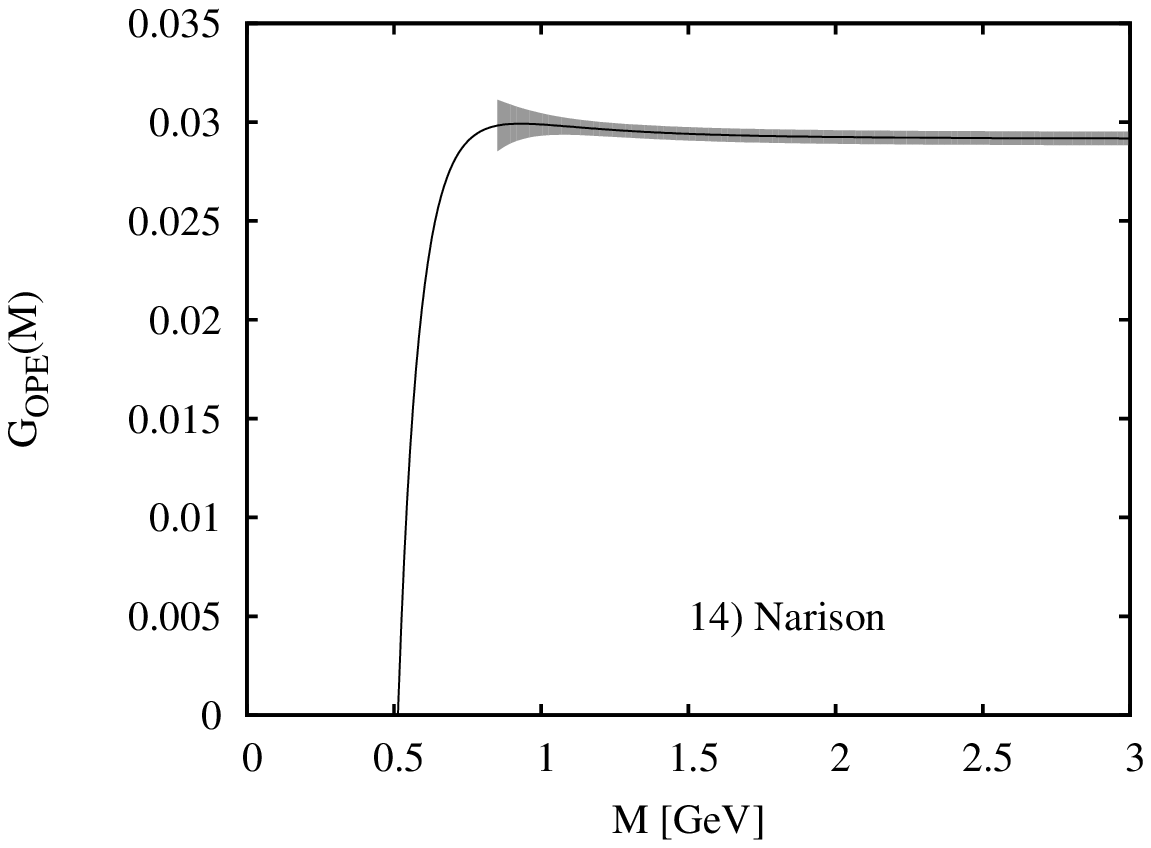} & \includegraphics[width=6.7cm,clip]{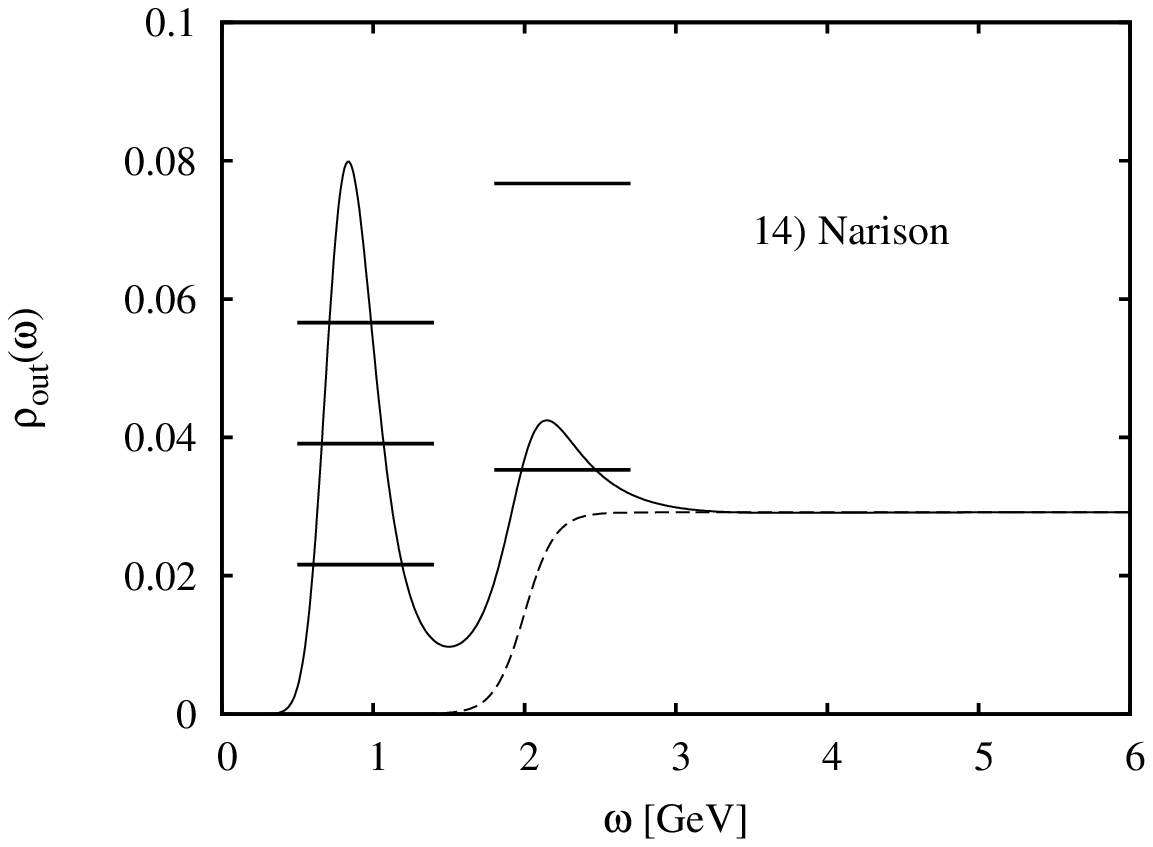} \\
\includegraphics[width=6.7cm,clip]{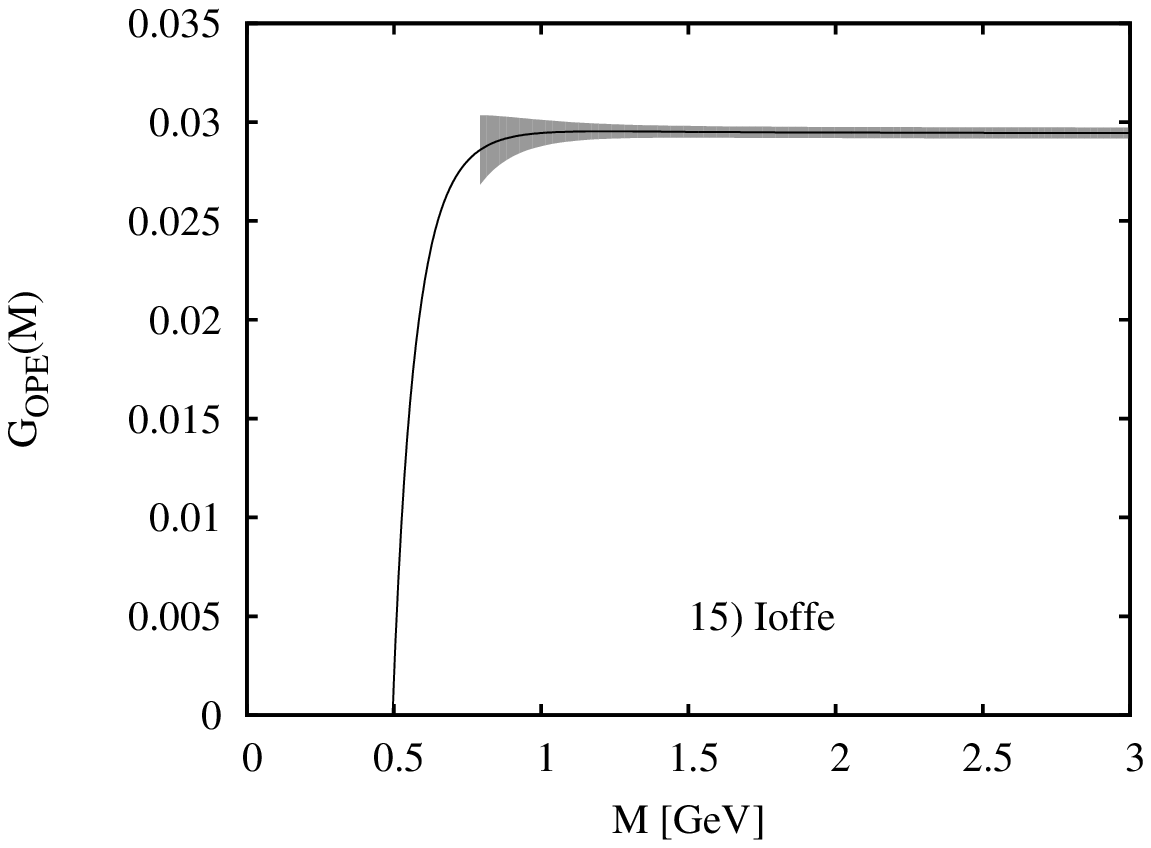} & \includegraphics[width=6.7cm,clip]{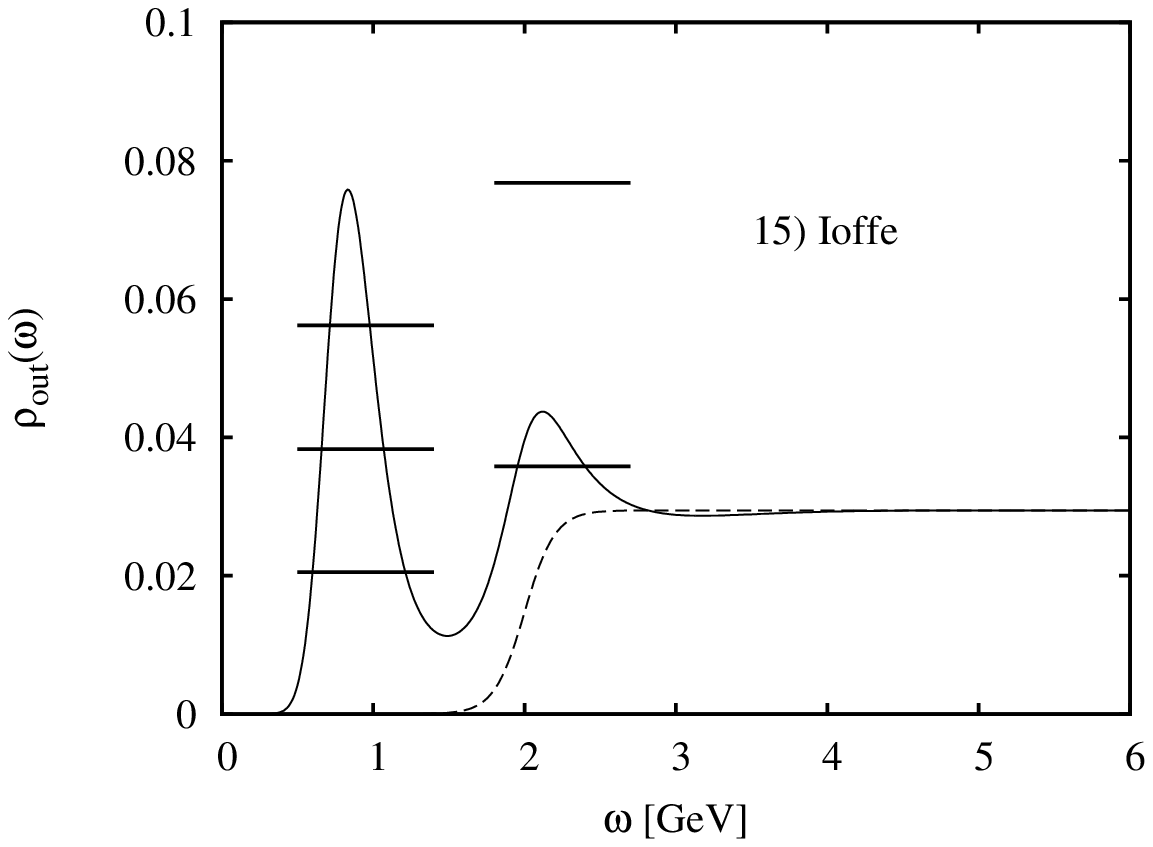}
\end{tabular}
\caption{On the left, the central values of $G_{OPE}(M)$ with the errors $\sigma(M)$ are given. 
The lower boundary of the shown errors corresponds to $M_{\mathrm{min}}$, below which 
the OPE does not converge. These plots should be compared with Fig. \ref{fig:mockdata}. 
On the right, the results of the MEM analysis using these OPE data are displayed. 
The dashed lines show the default model, and 
the horizontal bars stand for the values of the spectral 
function, averaged over the peaks $\langle \rho_{\mathrm{out}} \rangle_{\omega_1 ,\omega_2}$ and the 
corresponding ranges 
$\langle \rho_{\mathrm{out}} \rangle_{\omega_1 ,\omega_2} \pm \langle \delta \rho_{\mathrm{out}} \rangle_{\omega_1 ,\omega_2}$. 
Their extent shows the averaged interval $(\omega_1, \omega_2)$. 
For the lower two figures the lower error bars 
of the second peak are not shown because they lie below $\rho(\omega) = 0$.}
\label{fig:Opethreesets}
\end{figure}

Having finished all the necessary preparations, we now proceed to the actual MEM analysis of the real OPE 
data. First, we show the central values of the right-hand side of Eq. (\ref{eq:operesult}) and the 
corresponding errors for the three parameter sets of Refs. \citen{Colangelo,Narison,Ioffe} on the left 
side of Fig. \ref{fig:Opethreesets}. 
Comparing these figures with Fig. \ref{fig:mockdata}, we see that 
the OPE results and the mock data obtained from Eq. (\ref{eq:spectralfunc})
exhibit a very similar behavior, even in the region smaller than $M_{\mathrm{min}}$, 
below which we have no control over the convergence of the OPE.

Using these data, we have carried out the MEM analysis. For the default model, we have adapted 
the function 
\begin{equation}
m(\omega) = \frac{1}{4\pi^2}\Bigl( 1 + \eta(\alpha_s) \Bigl)\frac{1}{1+ e^{(\omega_0 -\omega) / \delta}}
\end{equation}
with parameters $\omega_0 = 2.0$ GeV and $\delta = 0.1$ GeV, which we have found in our investigation of 
the mock data to give the best reproduction of the $\rho$-meson peak with the smallest 
relative error and no large artificial peaks. 
The results are shown on the right side of Fig. \ref{fig:Opethreesets}. 
We clearly see that all three data sets give a significant lowest peak, which 
corresponds to the $\rho$-meson resonance. To determine the peak position, we 
simply adopt the value, where the peak reaches its highest point. The uncertainty of 
this quantity has already been estimated in our mock-data analysis and we employ 
the number that has been obtained there to specify the error of our 
final results that are given in the first line of Table \ref{finalresults}.
\begin{table}[t]
\begin{center}
\caption{Final results for the three parameter sets. The corresponding errors were 
determined from our analysis of mock data in the previous section.} 
\label{finalresults}
\begin{tabular}{lcccc} \hline
 & Colangelo and Khodjamirian \cite{Colangelo} & \hspace*{0.5cm} Narison \cite{Narison} \hspace*{0.5cm}& Ioffe \cite{Ioffe} & Experiment\\ \hline
$m_{\rho}$ & $0.75 \pm 0.06 \,\,\mathrm{GeV}$ & $0.84 \pm 0.10 \,\,\mathrm{GeV}$ 
& $0.83 \pm 0.09 \,\,\mathrm{GeV}$ & $0.77  \,\,\mathrm{GeV}$  \\ 
$F_{\rho}$ & $0.172 \pm 0.038\,\,\mathrm{GeV}$ & $0.190 \pm 0.049\,\,\mathrm{GeV}$ 
& $0.186 \pm 0.049\,\,\mathrm{GeV}$ &  $0.141 \,\,\mathrm{GeV}$ \\ \hline
\end{tabular}
\end{center}
\end{table}   

Next, fitting the spectral functions of Fig. \ref{fig:Opethreesets} to 
a relativistic Breit-Wigner peak plus a second-order polynomial background, 
as was done in Fig. \ref{fig:fitresult}, we have determined the 
coupling strength $F_{\rho}$ from our obtained spectral function, 
leading to the results given in the second line of Table \ref{finalresults}. 
As could be expected from 
our experience with the mock data, we get results that are 
all somewhat larger than the experimental value.

\subsubsection{Dependence of the $\rho$-meson peak on the values of the condensates}
Looking at Tables \ref{parameters} and \ref{finalresults}, it is interesting to observe that even though 
the parameter sets of Refs. \citen{Narison} and \citen{Ioffe} are quite different, they lead 
to almost identical results. This fact can be explained from the dependences of the 
properties of the $\rho$-meson resonance on $\langle \bar{q} q \rangle$, 
$\big \langle \frac{\alpha_s}{\pi} G^2 \big \rangle$ and $\kappa$, 
as will be shown in this subsection. 

\begin{figure}[t]
\centering
\includegraphics[width=6.7cm,clip]{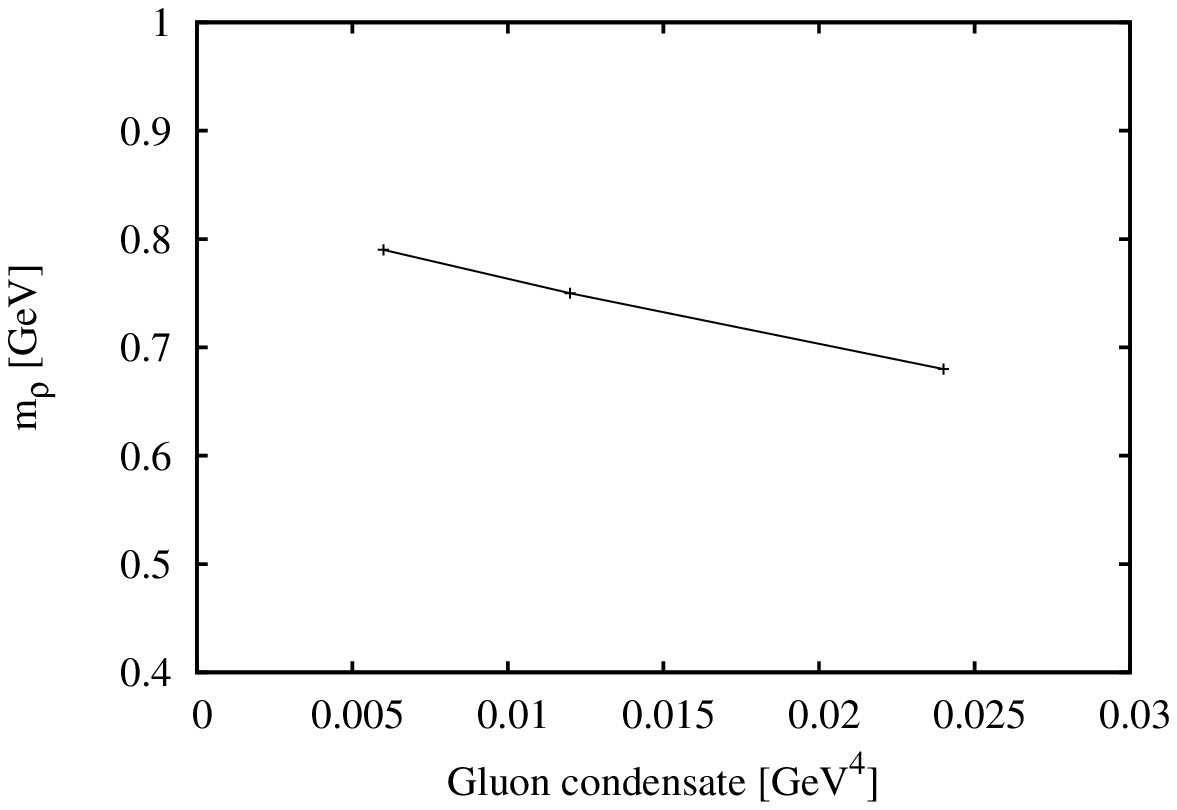}
\includegraphics[width=6.7cm,clip]{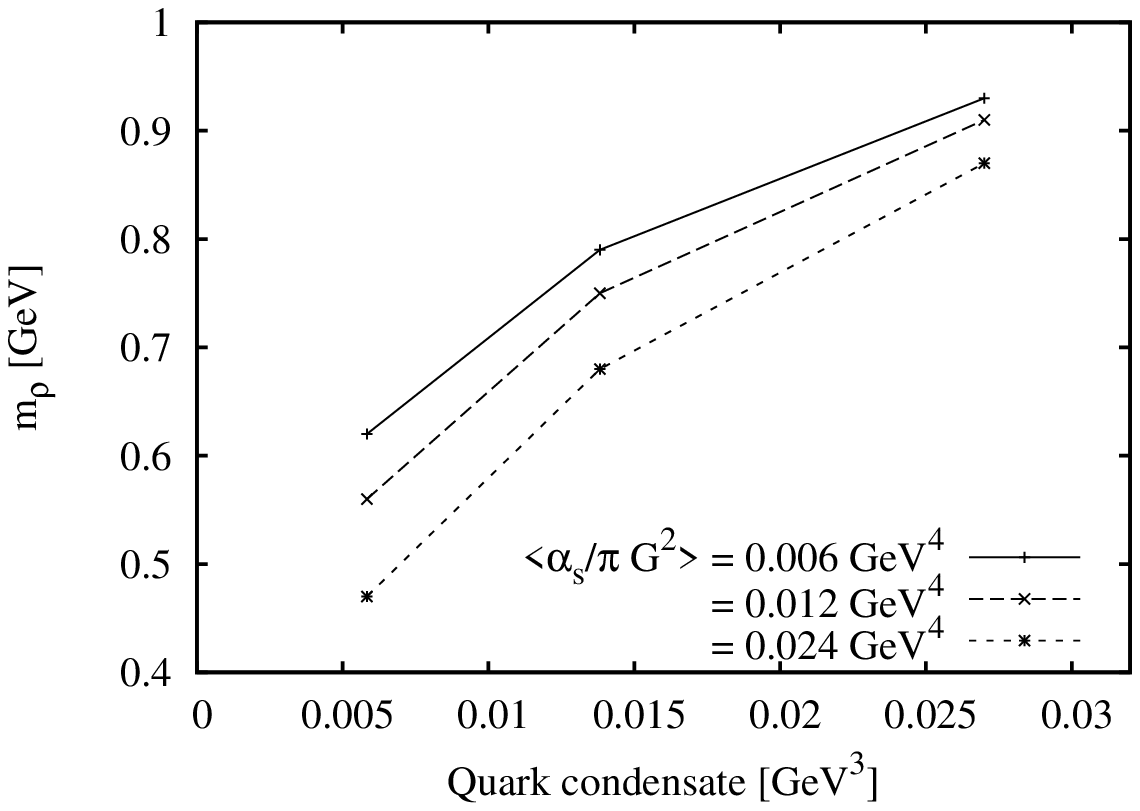}
\caption{In the left figure, the mass $m_{\rho}$ is shown as a function of the gluon condensate 
$\langle \frac{\alpha_s}{\pi} G^2 \rangle$. 
Here, the value of $\langle \bar{q} q \rangle = -(0.24)^3\,\mathrm{GeV}^3$ has been used 
for the quark condensate. 
In the right figure, $m_{\rho}$ is shown as a function of the quark condensate $\langle \bar{q} q \rangle$ for 
three different values of the gluon condensate. To obtain these plots, the errors of Ref. \citen{Colangelo} 
have been used.}
\label{fig:dependence.glu.qbarq}
\end{figure}

Investigating the relation between the $\rho$-meson resonance and the condensates is 
also interesting in view of the behavior of this hadron at finite temperature or 
density, as the values of the condensates will change in such environments. 
This will in turn alter the QCD sum rule predictions for the various hadron 
properties. 
A detailed study of this kind of behavior of the $\rho$-meson is nevertheless beyond the scope of 
the present paper and is left for future investigations. Here, we only discuss the change in the 
mass of the $\rho$-meson when the values of the condensates are modified by 
hand. The behavior of the peak position $m_{\rho}$ is shown in Fig. \ref{fig:dependence.glu.qbarq}. 
For obtaining these results, we have used the errors of Ref. \citen{Colangelo} and the 
corresponding values of $\kappa$ and $\alpha_s$, but have confirmed that the qualitative 
features of Fig. \ref{fig:dependence.glu.qbarq} do not depend on the explicit values 
of those parameters. 
It is seen that while $m_{\rho}$ decreases somewhat when $\langle \frac{\alpha_s}{\pi} G^2 \rangle$ 
increases, its value grows quite strongly when $\langle \bar{q} q \rangle$ increases, irrespective 
of the value of the gluon condensate. 
We found that the coupling strength $F_{\rho}$ shows the same qualitative behavior, slightly 
decreasing with increasing $\langle \frac{\alpha_s}{\pi} G^2 \rangle$, and strongly 
increasing with increasing $\langle \bar{q} q \rangle$. 
A similar tendency for $m_{\rho}$ and $F_{\rho}$ has also been observed 
in Ref. \citen{Leinweber}. 
This result shows that the quark condensate plays an essential role in determining the properties 
of the $\rho$-meson. 

It is important to note here that the correlation between $m_{\rho}$ and $\langle \bar{q} q \rangle$ 
to a large part occurs due to the last term in $G_{OPE}(M)$ of Eq. (\ref{eq:operesult}), which contains 
the squared quark condensate. This means that a similar (but weaker) correlation exists between 
$m_{\rho}$ and $\kappa$, which is also present in the last term of $G_{OPE}(M)$. 
Hence, we can now understand why the parameters of Refs. \citen{Narison} and 
\citen{Ioffe} give such similar results: while the large value of the gluon condensate of 
Ref. \citen{Narison} should lead to a smaller $m_{\rho}$, this effect is compensated by the large 
value of $\kappa$, which shifts the mass upwards. Therefore, the sum of these changes 
cancel each other out to a large degree, the net effect being almost no change in the spectral 
function for both cases.  

\section{Summary and conclusion}
We have applied the MEM technique to the problem of analyzing QCD sum rules. 
Using MEM has the advantage that we are not forced to introduce 
any explicit functional form 
of the spectral function, such as the ``pole + continuum" ansatz that 
has often been employed in QCD sum rule studies. This therefore allows us to investigate any 
spectral function without prejudice to its actual form. Furthermore, with this 
technique, we have direct access to the spectral function without the need for interpreting 
quantities that depend on unphysical parameters such as the Borel mass $M$ and the 
threshold parameter $s_{\mathrm{th}}$. 

To check whether it is really possible to apply MEM to QCD sum rules, we have investigated 
the vector meson channel in detail, first with mock data obtained from a realistic model 
spectral function, and then with the actual Borel-transformed results of the operator 
product expansion. The main results of this investigation are as follows: 
\begin{itemize}
\item Most importantly, demonstrating that it is possible to extract a significant peak in the spectral 
function, which corresponds to the $\rho$-meson resonance, 
we could show that the MEM technique is quite useful for analyzing QCD 
sum rules. 
For both mock or OPE data, we were able to reproduce the experimental $\rho$-meson mass 
$m_{\rho}$ with a precision of about 10\% and the coupling strength $F_{\rho}$ with a precision 
of about 30\%.
\item We have found that owing to the properties of the kernel of Eq. (\ref{eq:kernel}) appearing in 
QCD sum rules, the 
default model $m(\omega)$ has to be chosen according to the correct behavior of the spectral function 
at low energy. We therefore have taken a default model that tends to zero 
at $\omega = 0$. 
On the other hand, to give the correct behavior at large energies, the same default model is constructed 
to approach the perturbative value in the high-energy region.
\item The position of the $\rho$-meson peak in the spectral function 
has turned out to be quite stable against changes 
in various parameters of the calculation, such as the details of the default model or the 
range of the analyzed Borel mass region. We have shown that changing these parameters leads 
to a fluctuation of the peak position of only 20 - 50 MeV. 
\item Concerning the width of the lowest lying peak, 
we are unable to reproduce the value of the input spectral function 
of the mock data with any reasonable precision and have shown that the reason for this difficulty 
comes from the insufficient sensitivity of the data $G_{\rho}(M)$ on the detailed form of the 
$\rho$-meson peak. To accurately estimate the width, one needs not only to go 
to higher orders in the OPE, but also has to have much more precise information on the values 
of the condensates than is available today. 
\item Accompanied by a steep rise in the default model $m(\omega)$, we have observed the 
appearance of artificial peaks in the output spectral function of the MEM analysis. 
These peaks are MEM artifacts and one has to be careful not 
to confuse them with the actual peaks that are predicted using the data $G_{OPE}(M)$.     
\end{itemize} 
These results are promising and encourage us to apply this approach to other 
channels, including baryonic ones. 
It would also be interesting to apply this method to 
the behavior of hadrons at finite density or temperature, as it would 
become possible to directly observe the change in the spectral function in hot or 
dense environments by this approach. 
Furthermore, it would be of interest to study the various exotic channels, 
containing more than three quarks. In these channels, the scattering states presumably 
play an important role and the bayesian approach of this paper could help 
clarify the situation and provide a natural way of distinguishing genuine resonances 
from mere scattering states.   

Even though these are interesting subjects for future studies, we want to emphasize here 
that the uncertainties involved for each channel can differ considerably and one thus 
should always carry out a detailed analysis with mock data for each case, before 
investigating the actual sum rules. 
This procedure is necessary to check whether 
it is possible to obtain meaningful results from the 
MEM analysis of QCD sum rules.

\section*{Acknowledgements}
This work was partially supported by KAKENHI under Contract Nos.19540275  and 22105503. 
P.G. gratefully acknowledges the support by the Japan Society for the 
Promotion of Science for Young Scientists (Contract No.21.8079). 
The numerical calculations of this study have been partly performed on the super grid 
computing system TSUBAME at Tokyo Institute of Technology.

%

\end{document}